\documentclass{emulateapj}

\newcommand{\lya}{Ly$\alpha$}

\def\spose#1{\hbox to 0pt{#1\hss}}

\def\lya{\ifmmode {\rm\,Ly\alpha}\else ${\rm\,Ly\alpha}$\fi}
\def\Mdot {\ifmmode {\rm {\dot M}} \else ${\rm {\dot M}}$\fi}

\def\kms{\ifmmode {\rm\,km\,s^{-1}}\else
    ${\rm\,km\,s^{-1}}$\fi}
\def\kmsMpc{\ifmmode {\rm\,km\,s^{-1}\,Mpc^{-1}}\else
    ${\rm\,km\,s^{-1}\,Mpc^{-1}}$\fi}

\def\msun{\ifmmode {\rm\,M_\odot}\else ${\rm\,M_\odot}$\fi}
\def\Msun{\ifmmode {\rm\,M_\odot}\else ${\rm\,M_\odot}$\fi}
\def\lsun{\ifmmode {\rm\,L_\odot}\else ${\rm\,L_\odot}$\fi}
\def\Lsun{\ifmmode {\rm\,L_\odot}\else ${\rm\,L_\odot}$\fi}
\def\rsun{\ifmmode {\rm\,R_\odot}\else ${\rm\,R_\odot}$\fi}
\def\Rsun{\ifmmode {\rm\,R_\odot}\else ${\rm\,R_\odot}$\fi}

\def\cm{{\rm\,cm}}
\def\cm3{\ifmmode {\rm\,cm^{-3}}\else ${\rm\,cm^{-3}}$\fi}

\def\ergps{\ifmmode {\rm\,erg\,s^{-1}}\else ${\rm\,erg\,s^{-1}}$\fi}
\def\ergpscm2{\ifmmode {\rm\,erg\,s^{-1}\,cm^{-2}}\else
    ${\rm\,erg\,s^{-1}\,cm^{-2}}$\fi}

\def\deg{\ifmmode {^{\circ}}\else {$^\circ$}\fi}
\def\degr{\ifmmode {^{\circ}}\else {$^\circ$}\fi}
\def\degs{\ifmmode {^{\circ}}\else {$^\circ$}\fi}

\def\etal{{et al.~}}

\def\h3Mpc{h^{-3}{\rm Mpc}^3}
\def\Ho{\ifmmode {\rm\,H_0}\else ${\rm\,H_0}$\fi}
\def\hnot{\ifmmode {\rm\,H_0}\else ${\rm\,H_0}$\fi}
\def\h0{\ifmmode {\rm\,H_0}\else ${\rm\,H_0}$\fi}
\def\hnotunit{\ifmmode {\rm\,km\,s^{-1}\,Mpc^{-1}}\else
    ${\rm\,km\,s^{-1}\,Mpc^{-1}}$\fi}
\def\qnot{\ifmmode {\rm\,q_0}\else ${\rm q_0}$\fi}
\def\q0{\ifmmode {\rm\,q_0}\else ${\rm q_0}$\fi}

\def\mic{\ifmmode {\rm\,\mu m}\else ${\rm \mu m}$\fi}
\def\micron{\ifmmode {\rm\,\mu m}\else ${\rm \mu m}$\fi}
\def\microns{\ifmmode {\rm\,\mu m}\else ${\rm \mu m}$\fi}

\def\arcsec{\ifmmode {^{\prime\prime}}\else $^{\prime\prime}$\fi}
\def\asec{\ifmmode {^{\prime\prime}}\else $^{\prime\prime}$\fi}
\def\arcmin{\ifmmode {^{\prime}}\else $^{\prime}$\fi}
\def\amin{\ifmmode {^{\prime}}\else $^{\prime}$\fi}

\def\secper{\ifmmode \rlap.{^{s}}\else $\rlap{.}{^{s}} $\fi}
\def\minper{\ifmmode \rlap.{^{m}}\else $\rlap{.}{^m} $\fi}
\def\magper{\ifmmode \rlap.{^{m}}\else $\rlap{.}{^m} $\fi}
\def\farcs{\ifmmode \rlap.{^{\prime\prime}}\else
    $\rlap.{^{\prime\prime}}$\fi}
\def\arcsper{\ifmmode \rlap.{^{\prime\prime}}\else
    $\rlap.{^{\prime\prime}}$\fi}
\def\arcmper{\ifmmode \rlap.{^{\prime}}\else
    $\rlap.{^{\prime}}$\fi}
\def\spose#1{\hbox to 0pt{#1\hss}}
\def\simlt{\mathrel{\spose{\lower 3pt\hbox{$\mathchar"218$}}
     \raise 2.0pt\hbox{$\mathchar"13C$}}}
\def\simgt{\mathrel{\spose{\lower 3pt\hbox{$\mathchar"218$}}
     \raise 2.0pt\hbox{$\mathchar"13E$}}}

\slugcomment{Draft Version 2008 November 13}

\shorttitle{}
\shortauthors{Zirm et al.}

\begin{document}

\title{Evolution of the Lyman-$\alpha$ Halos around High-Redshift Radio Galaxies}

\author{Andrew W. Zirm}
\affil{The Johns Hopkins University, Baltimore, MD 21218}
\email{azirm@pha.jhu.edu}

\author{Arjun Dey}
\affil{National Optical Astronomy Observatory, Tucson, AZ 85719}
\email{dey@noao.edu}

\author{Mark Dickinson}
\affil{National Optical Astronomy Observatory, Tucson, AZ 85719}
\email{med@noao.edu}

\and

\author{Colin J. Norman}
\affil{The Johns Hopkins University, Baltimore, MD 21218}
\email{norman@stsci.edu}

\begin{abstract}

  We have obtained the first constraints on extended Ly$\alpha$
  emission at $z \sim 1$ in a sample of five radio galaxies. We detect
  Ly$\alpha$ emission from four of the five galaxies.  The Ly$\alpha$
  luminosities range from $0.1 - 4 \times 10^{43}$ erg s$^{-1}$ and
  are much smaller than those observed for halos around higher
  redshift radio galaxies.  If the $z\approx1$ radio galaxies are the
  descendents the $z\simgt2$ radio galaxies, then their Ly$\alpha$
  luminosities evolve strongly with redshift as $\sim
  (1+z)^{5}$. There do not appear to be strong correlations between
  other parameters, such as radio power, suggesting that this observed
  evolution is real and not an observational artifact or secondary
  correlation.  We speculate that this evolution of luminous halos may
  be due to gas depletion (as gas cools, settles, and forms stars)
  accompanied by an overall rise in the mean gas temperature and a
  decrease in specifc star-formation rate in and around these massive
  galaxies.

\end{abstract}


\section{Introduction\label{sec:intro}}

The bright end of the luminosity function is dominated today by giant
elliptical and cD galaxies.  The history of these massive galaxies
traces the evolution of the highest peaks in the initial density
perturbation spectrum, and thus potentially places strong constraints
on paradigms of structure formation.  The host galaxies of
high-redshift radio galaxies ($z \simgt 1$; HzRGs) are likely the
progenitors of these modern-day giants: the HzRG $K$--band Hubble
diagram is well--fit by the `passive' evolution of a stellar
population with a high formation redshift \citep*[e.g.,][]{McCarthy93,
  DeBreucketal02}, and NICMOS continuum images of $0.8 < z < 1.8$ 3CR
radio galaxies show that the starlight distributions are round and
symmetric, mostly with $R^{1/4}$--law light profiles
\citep*{Zirmetal99, ZirmDickinsonDey03}.  Furthermore, the inferred
stellar masses derived from modelling of the spectral energy
distributions (SEDs), including data from the {\it Spitzer Space
  Telescope}, define the upper envelope of the mass function for all
high redshift galaxies \citep*{Seymouretal07}.  In addition, age
dating studies of a few $z\sim 1-1.5$ HzRGs suggests that they formed
the bulk of their stars at significantly higher redshifts, $z \gg 2$,
and evolved fairly `passively' thereafter
\citep*{StocktonKelloggRidgway95, Dunlopetal96, Spinradetal97}.
Finally, radio galaxies at $z \simgt 2$ are generally found in
overdense, protocluster, environments \citep*[e.g.,][]{Venemansetal07,
  Kodamaetal07, Zirmetal08} again consistent with them marking high
density peaks in the dark matter field.

Many of the $z > 2$ radio galaxies, including those within
protocluster regions, are surrounded by giant ($\simgt 100$ kpc) \lya\
halos \citep*{McCarthyetal90, McCarthySpinradvanBreugel95, Deyetal97,
  Villar-Martinetal99, Reulandetal03}.  These halos have line
luminosities of several~$\times 10^{44}~{\rm erg\ s^{-1}}$, and
suggest the presence of huge gas reservoirs.  Initially these halos
were thought to be associated only with, and perhaps powered by, rare,
luminous radio galaxies.  However, the discovery by Steidel \etal
(2000)\nocite{Steideletal00} of two giant \lya\ ``blobs'' associated
with a galaxy overdensity at $z = 3.09$ and only loosely associated
with any galaxies with detectable UV continuum suggest that Ly$\alpha$
halos may be relevant to the formation of the most massive galaxies in
general.  Subsequent deep \lya\ imaging and spectroscopy of this same
protocluster field identified many lower luminosity, radio-quiet,
spatially-extended \lya\ emitters \citep*{Matsudaetal04,Saitoetal08}.

The origins of the \lya\ emitting gas remain ambiguous.  Monolithic
cooling of pristine, infalling gas would have an extremely short
lifetime ($\tau_{cool} \sim 10^{4}$ yrs; e.g., Dey et
al. 2005\nocite{Deyetal05}) and could not explain the prevalence of
the halo phenomenon.  However, if the ionized gas reservoirs were
replenished via further infall and/or outflows from the central source
the `monolithic' scenario may still be viable.  If we assume that the
halos are formed by a mixture of inflow and outflow, the morphology
and evolution of the \lya\ emission should provide clues to the nature
of that mix and by extension the galaxy formation process.  Another
possibility is that each Hydrogen atom is ionized more than once on
average.

Little is known about how these halos evolve with redshift.  As the
host galaxy and radio source age spectrally and dynamically, the
interaction between the halo and its embedded sources may alter in
form.  Between $z \sim 4$ and $z \sim 1$, the host galaxy is largely
assembled as a mature giant elliptical, and a hot young stellar
population is no more than a minor contributor to the total UV flux
\citep*[e.g.,][]{ZirmDickinsonDey03}.  The central ionization source
may shift from being dominated by young stars and starburst outflows
to being dominated by the interaction with the radio source and
photoionization by the central engine.  Understanding the connection
between young radio galaxies at $z \sim 3$ and the more mature objects
at $z \sim 1$ requires similar data at all redshifts, including
measurements of their emission line properties.  Moreover, studies of
the same emission lines at similar sensitivity can distinguish genuine
evolutionary trends from the consequences of line choice or
instrumental limitations.

\begin{deluxetable*}{lccccccc}
\tabletypesize{\scriptsize}
\tablecolumns{8}
\tablewidth{0in}
\tablecaption{Radio Galaxy Observations and Measurements\label{tab:obs}}
\tablehead{
\colhead{Galaxy} & \colhead{$z$} & \colhead{RA} & \colhead{Dec} & \colhead{NUV Prism} & \colhead{NUV Direct} & \colhead{Line Flux} & \colhead{Line Luminosity} \\
\colhead{} & \colhead{} & \colhead{(J2000)} & \colhead{(J2000)} & \colhead{Exp. Time (s)} & \colhead{Exp. Time (s)} & \colhead{$10^{-15}$ erg s$^{-1}$ cm$^{-2}$} & \colhead{$10^{42}$ erg s$^{-1}$}
}
\startdata
3C~210 & 1.17 & 08:58:09.9 & 27:50:52 & 2757 & 260 & $< 0.1$ & $< 1$ \\
3C~265 & 0.81 & 11:45:29.0 & 31:33:49 & 1330 & 360 & $28.6$ & $2.3$ \\
3C~266 & 1.28 & 11:45:43.4 & 49:46:08 & 5260 & 360 & $1.5$ & $36$ \\
3C~267 & 1.14 & 11:49:56.5 & 12:47:19 & 2709 & 240 & $0.2$ & $1.0$ \\
3C~324 & 1.21 & 15:49:48.9 & 21:25:38 & 3474 & 360 & $3.4$ & $20$ \\
\enddata
\end{deluxetable*}

In this paper, we present Ly$\alpha$ measurements of 5 radio galaxies
at $0.8<z<1.3$ obtained using the slitless prism mode with STIS on
HST.  This article is organized as follows: we present the
observations and data reduction in the next section
(\S~\ref{sec:obs}), the results in Section~\ref{sec:results}, a
discussion of these results in the context of theories of galaxy
formation in Section~\ref{sec:discussion} and finally a brief summary
of our conclusions in Section~\ref{sec:conclusions}.  Throughout this
paper we assume a cosmological model with $\Omega_{\Lambda} = 0.7$ and
$\Omega_{m} = 0.3$ and $H_{0} = 70$ km s$^{-1}$ Mpc$^{-1}$.

\section{Observations\label{sec:obs}}

We used the Space Telescope Imaging Spectrograph (STIS) on-board the
{\it Hubble Space Telescope} during 20 December 2001 - 12 April 2003
to image 5 high-redshift radio galaxies lying in the range $0.8<z<1.3$
(GO\#9166; PI A. Zirm). The target galaxies are selected from the 3CR
catalog \citep*{Bennett62} and all are known to have strong nebular
line emission based on their optical spectra.  At the redshift of the
target galaxies, the \lya\ line falls in the near-ultraviolet (NUV)
and thus requires space-based observation.  Observations were made
with the STIS NUV-MAMAs, ultraviolet sensitive array detectors, with
the prism, to obtain very low dispersion ($\sim$ 30\AA\ pixel$^{-1}$)
spectra over the $\lya$ line to produce, effectively, a line image.
The extremely dark UV sky background and absence of readout noise in
the STIS MAMAs enhance our sensitivity to low surface--brightness
emission.  Our resultant \lya\ images have depth comparable to those
of $z \sim 3$ objects in a few orbits, with much higher spatial
resolution.

The dispersion of the STIS prism mode at redshifted \lya\ at $z = 1$
is $\approx 3000$\kms~pixel$^{-1}$, sufficiently low that the
prism--dispersed \lya\ will essentially form a monochromatic emission
line image, with the galaxy continuum dispersed beneath it.  In
addition to the slitless prism data we also took direct images using
both the NUV-MAMA (which includes \lya) and the optical CCD to
determine the zeropoint for the source spectra.  The details of
exposure times are listed in Table~\ref{tab:obs}.

We extracted the Lyman-$\alpha$ fluxes in a simple manner.  Visual
inspection of the prism and direct image data show that the 2D spectra
are dominated by line emission.  The structures observed in the direct
image are reproduced when the prism is used rather than being smoothed
over larger areas.  3C~265 is the exception where both continuum and
the CIV emission line are also visible in the 2D spectrum.  However,
to model the spectra, even in the case of 3C~265, we have taken a
simple power-law continuum with both $f_{\lambda} \propto \lambda^{0}$
and $\propto \lambda^{-2}$ (i.e., flat in $f_{\lambda}$ and $f_{\nu}$
respectively).  We chose these two extrema based on the empirical
range of far-UV slopes observed for both star-forming galaxies
\citep*[e.g.,][]{MeurerHeckmanCalzetti99} and AGN
\citep*[e.g.,][]{VandenBerketal01, Telferetal02}.  We used the
tabulated dispersion relation for the prism and the direct NUV image
of each galaxy to construct continuum-only models of the prism
observations.  We thereby assumed that the morphology of the direct
image, which includes \lya\ , is representative of the spatial
distribution of continuum flux.  These models were scaled and
subtracted, to minimize the residuals, to leave only the line
emission.  We note that the exact details of the continuum-subtraction
should not significantly change the results.  Lyman-$\alpha$ fluxes
were then derived by performing photometry in apertures designed to
include the residual flux on these continuum-subtracted images and
masking the CIV flux in the case of 3C~265.  The resulting \lya\ line
fluxes luminosities are listed in Table~\ref{tab:obs}.

\begin{figure}[t]
\epsscale{1.0}
\plotone{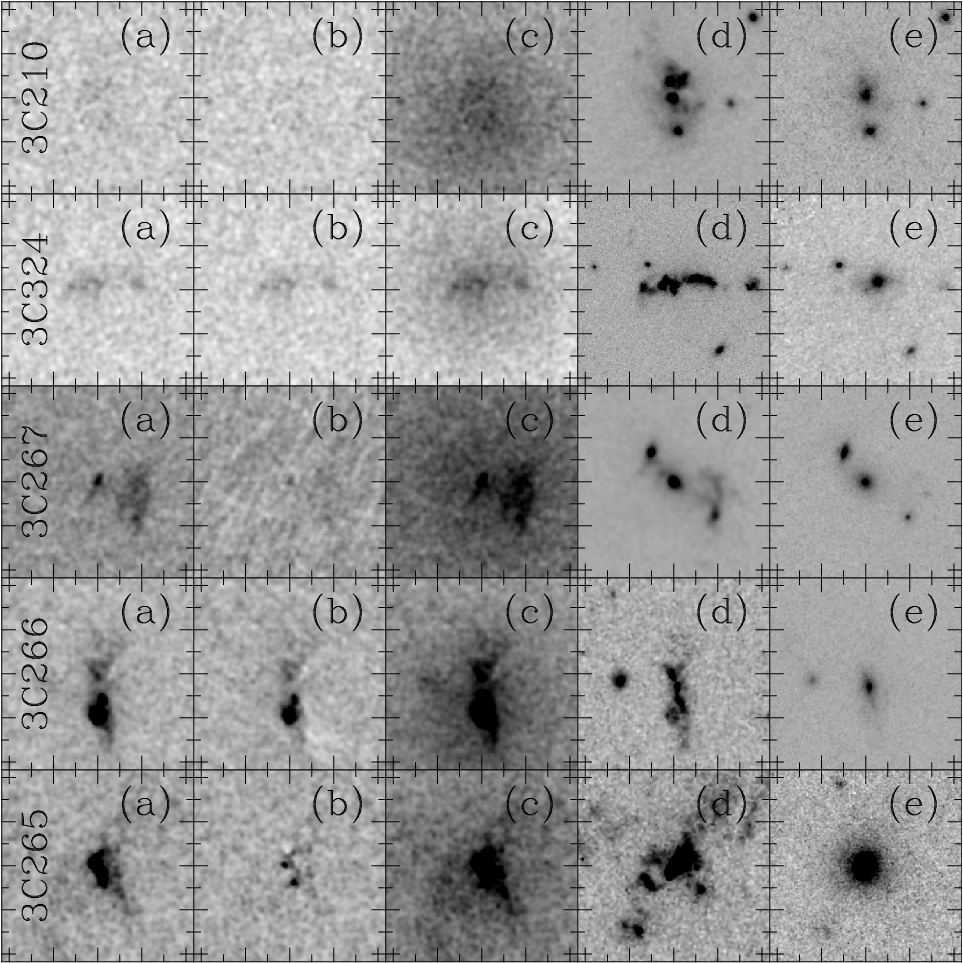}
\caption{2D prism spectrograms of the 5 radio galaxies (panel a) and
  their continuum-subtracted (panel b; Ly$\alpha$ only for all except
  3C~265, where CIV is also visible) data and how they would
  have appeared with a fiducial $L_{\rm Ly\alpha} = 1 \times 10^{44}$
  erg s$^{-1}$ halo (panel c). Panel d shows the observed optical image and panel e is the observed near-infrared image.  North is up, East to the left and the major tickmarks are separated by 2
  arcseconds.
  \label{fig:cuts}
}
\end{figure}

The raw prism data, the continuum subtracted data and the data with a
fiducial halo model added are presented in Figure~\ref{fig:cuts}.  The
fiducial model halo has $L_{\rm Ly\alpha} = 1 \times 10^{44}$ erg
s$^{-1}$ and follows a King model profile with core radius $= 10$ kpc,
similar to observations of \lya\ halos surrounding $z > 2$ radio
galaxies \citep*[e.g.,][]{Reulandetal03}.  The observed optical and
near-infrared data are from {\it HST} WFPC2 and NICMOS respectively.  

\section{Results\label{sec:results}}

\begin{figure}[b]
\epsscale{1.0}
\plotone{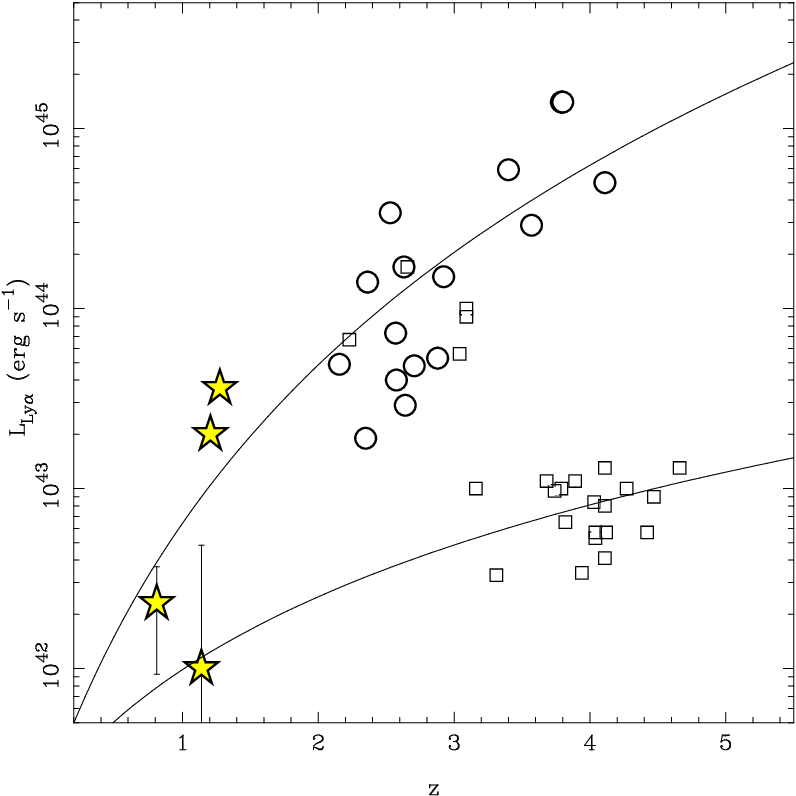}
\caption{Trend of \lya\ luminosity with redshift for the four radio
  galaxies presented here (stars), higher redshift radio galaxies
  (circles) and radio-quiet Ly$\alpha$ halos
  \citep*[squares;][]{Steideletal00, Matsudaetal04, Deyetal05,
    Saitoetal08}.  The two solid lines are power-law evolution tracks
  of $(1+z)^{5}$ and $(1+z)^{2.3}$ to guide the eye.\label{fig:evol}}
\end{figure}

Of the five $z \sim 1$ radio galaxies targeted, four have significant
or marginal detections of Lyman-$\alpha$ flux. 3C~210 shows no line
emission to a limit of $L_{\rm Ly\alpha} \simgt 10^{42}$ erg s$^{-1}$.
Of the four with \lya\ , two are bright, 3Cs 266 and 324 with $L_{\rm
  Ly\alpha} = 3.6$ and $2.0 \times 10^{43}$ erg s$^{-1}$ respectively,
while 3Cs 267 and 265 are both a factor of ten less luminous (see
Table~\ref{tab:obs}).  The evolution of the \lya\ luminosity around
powerful radio galaxies with redshift is shown in
Figure~\ref{fig:evol}.  The stars are the galaxies from the current
study while the circles are higher redshift radio galaxies
\citep*{McCarthyetal90, McCarthySpinradvanBreugel95, Deyetal97,
  Reulandetal03, Villar-Martinetal03, Zirmetal05, Venemansetal07,
  Villar-Martinetal07} and the squares are radio-quiet extended \lya\
emitters \citep*[`blobs';][]{Steideletal00, Matsudaetal04}.  The solid
lines are `by eye' power law evolution models with $L_{\rm Ly\alpha}
\propto (1+z)^{\alpha}$ where $\alpha = 5.0$ and $2.3$.  The radial
extent of these $z \sim 1$ halos are about a factor of two smaller
than the high-redshift halos.  There is no similarly strong redshift
correlation with radio power or stellar mass of the radio host for
HzRGs in general (see Seymour et al. 2007\nocite{Seymouretal07}).  In
any case, it is clear that the $100$kpc, $10^{44}$ erg s$^{-1}$ halos
generally seen around powerful radio galaxies at $z \simgt 2$ are not
present around their counterparts (which are not necessarily their
descendents) at $z \sim 1$.

\begin{figure}[b]
\epsscale{0.45}
\plotone{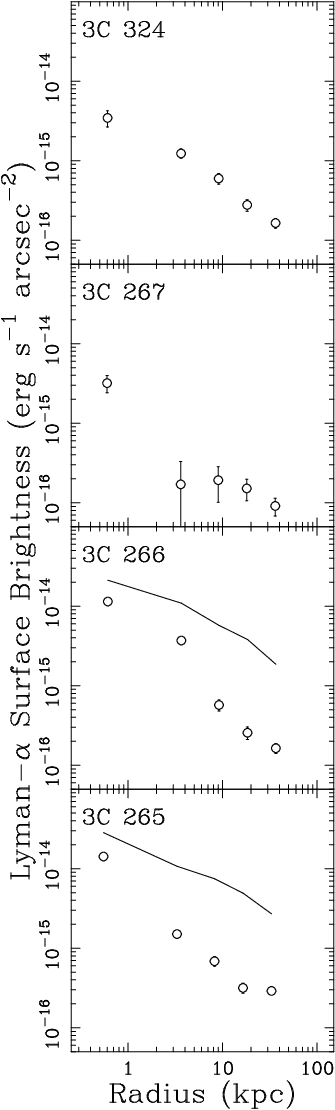}
\caption{Radial profiles of the Ly$\alpha$ emission around the 4 radio
  galaxies with significant line flux. For 3C~265 the regions where
  CIV emission is seen were masked for these surface-brightness
  measurements.  The solid lines in the 3C~265 and 3C~266 panels
  represent the surface-brightness profiles for the fiducial model
  halo ($L_{\rm Ly\alpha} = 10^{44}$ erg s$^{-1}$). \label{fig:radial}}
\end{figure}

At least two of the radio galaxies (3C~265 and 266) show Ly$\alpha$
emission with significant angular extent (Figure~\ref{fig:radial}).
For these galaxies, we also plot the fiducial halo model in the same
apertures for comparison.  In addition to the difference in overall
luminosity, the lower redshift halos also appear to have steeper
surface-brightness profiles with smaller extent along with some
constant surface-brightness emission associated with the UV continuum.
The sizes of the detectable line-emitting regions are only tens of
kpcs as opposed to the higher-redshift examples with radii close to
100 kpc.  More specifically, for the well-detected case of 3C~265, the
morphology of the extended line emission seems rather different from
the halos seen at high redshift.  For the other galaxies, our data are
not deep enough to reveal much morphological information.  For 3C~265,
however, we can see that the extended line emission (panel a in
Fig.~\ref{fig:cuts}) roughly follows the spatial distribution of the
UV continuum (panel d) which is aligned with the double-lobed radio
axis, perhaps indicating that this line emission is excited by the
AGN.  This in contrast to the higher redshift halos which generally
show little or no UV continuum associated with the line emission.

\begin{figure}
\epsscale{1.0}
\plotone{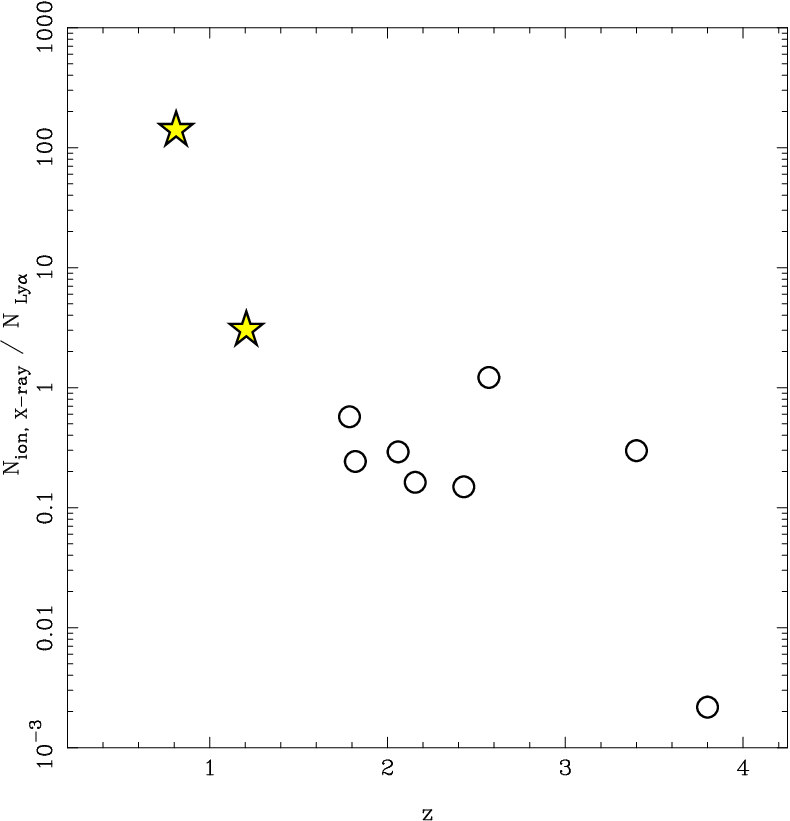}
\caption{Ratio of number of ionizing photons as inferred from the
  X-ray luminosity to the number inferred from the Ly-$\alpha$
  luminosity versus redshift for a sample of radio galaxies drawn from
  the literature (circles) and from the current work
  (stars).\label{fig:Xray}}
\end{figure}

To investigate the energetics of the halos, we have used X-ray core
measurements from the literature \citep*{CrawfordFabian96,
  HardcastleWorrall99, Fabianetal02, Pentericcietal02, Derryetal03,
  Fabianetal03, Scharfetal03, Overzieretal05} and Ly-$\alpha$
luminosities to make two different estimates of the UV ionizing
radiation in the radio galaxy environments.  We assume a single
ionizing photon results in a single Ly$\alpha$ photon.  For the X-ray
data we assume a single power-law spectrum from the rest-frame X-ray
to the UV to estimate the number of ionizing photons due to the AGN.
In Figure~\ref{fig:Xray} we show the ratio of the two ionizing
luminosities versus redshift.  While this plot is a bit sparse and
consists of rather heterogeneous datasets, it seems that there may be
a trend for the high redshift halos to require a power source beyond
the X-ray luminosity while at $z \sim 1$ the X-rays may be sufficient
to power the extended line emission.  We discuss the implications of
this result further in the following section.

\section{Discussion\label{sec:discussion}}

We have presented slitless NUV spectroscopy of five $z \sim 1$
powerful radio galaxies.  These spectra show for the first time the
considerable redshift evolution of the extended \lya\ luminosity
surrounding powerful radio galaxies.  The morphology of the extended
line emission also seems to change from a centrally-concentrated halo
at high redshift to a more evenly distributed surface-brightness
profile at lower redshift associated with the known (extended) UV
continuum.  The size of the halos are also about a factor of two (at
minimum) smaller than the high-redshift examples.  These changes
together suggest that we may be seeing a change in the dominant
process responsible for the line emission as a function of redshift.
Furthermore, the lack of similar halos around comparable power radio
galaxies, along with the discoveries of radio-quiet \lya\ halos,
implies that halos are not solely a feature of radio galaxies but may
be associated more generally with galaxy and structure formation.

By making observations of \lya\ emission over a range of redshifts we
study the evolution of the gaseous environments of massive galaxies
where other methods, such as X-ray imaging, are impractical at higher
redshifts.  The luminosity and structure of the line emitting gas
provides clues to the dynamical and excitation states of the gas.  We
can use simulations to put these observations in context.  Dijkstra,
Haiman and Spaans (2006a,b) \nocite{DijkstraHaimanSpaans06I,
  DijkstraHaimanSpaans06II} have quantified the relationship between
different infall and cooling scenarios and \lya\ surface brightness
using a set of simulations.  Their range of initial conditions and
assumptions span the gap between two extremes.  First, that the
cooling timescale is short and tracks the loss of gravitational
potential energy as the gas falls into the center after being shock
heated at the virial radius \citep*[e.g.,][]{HaimanSpaansQuataert00}.
Second, that the gas is accreted cold (i.e., no virial shock heating)
and only subsequently its kinetic energy is converted to thermal
energy resulting in \lya\ emission \citep*[e.g.,][]{DekelBirnboim06}.
These simulations confirm and quantify the intuitive view that the
line emission is more centrally-concentrated for the case where
heating and cooling only occur once the gas has collapsed to the
center and that more extended emission is seen where clumps of gas are
assumed to cool as they descend into the gravitational potential.  It
should be noted, however, that both scenarios produce extended halos
of line emission.

In addition to either or both of these infall scenarios the gas may
also be ionized by UV-luminous young and forming stars and by the
central AGN.  
For 3C~265, the close correlation between the UV continuum emitting
regions and the \lya\ emission suggests that the ionizing source is
low-level star-formation, young stars and photoionization by AGN light
in these areas.  Our observations indicate that the halo phenomenon as
seen at high redshift has changed significantly by $z \sim 1$.
This inference is further supported by our analysis of the energetics
of the radio galaxies over the same range of redshift.  While at high
redshift $z > 2$ the virial shocking of infalling gas may dominate the
\lya\ emission it seems that by $z \sim 1$ the X-ray output of the
central AGN may be sufficient to ionize the surrounding gas.

Under any scenario, it is clear that the gaseous environments of
massive galaxies are subject to several possible changes during the
more than three billion years between $z \sim 2.5$ and $z \sim 1$.
Gas is consumed in star-formation and black hole accretion, it is
enriched with metals and deposited back in the IGM via supernovae and
AGN-driven outflows and the IGM itself is disrupted and shock-heated
via galaxy interactions and mergers.  Overall, however, the amount of
cool gas surrounding galaxies decreases with cosmic time while
elliptical galaxies and galaxy clusters retain vast reservoirs of hot
gas.  If we associate the high redshift, luminous halos with the
initial accretion and starburst, the faint halos at later times
suggests that the gas has either been depleted via star formation and
accretion or that the cooling time has increased sufficiently to
quench infall.  Cooling radiation and subsequent star-formation
dominates the halo luminosity at high redshift with some additional
contribution associated with the AGN.  As the gas cools and forms
stars the bulk of the line luminosity disappears with only the
residual, AGN and star-formation related emission remaining by $z \sim
1$.  This residual emission is over a smaller area (radius $\simlt
50$kpc) and is associated with the UV continuum emission.  

\section{Conclusions\label{sec:conclusions}}

We have used STIS NUV--MAMA slitless spectroscopy of five $z\sim 1$
powerful radio galaxies to determine their \lya\ properties and to
specifically look for luminous, extended halos of line emission as are
seen around $z \simgt 2$ radio galaxies.  While we find that at least
two of the targets have extended emission line regions it is unclear
from these data alone whether this \lya\ is directly related to the
high-$z$ halos.  The emission line morphologies are rather different
and the luminosities are an order-of-magnitude lower than at higher
redshift.  The physical extent of the halos are smaller by about a
factor of two compared to the $z \simgt 2$ examples.  We have argued
that based on the available X-ray data it is likely that the dominant
mechanism for producing \lya\ photons has changed from one related to
gas infall and cooling at high redshift to one dominated by
photo-ionization by young stars and the AGN by $z \sim 1$.

Based on these data it is clear that \lya\ studies of massive galaxies
over a range of redshifts will provide clues to the state of the
gaseous environment and the history of galaxy feedback in the local
IGM.  In the scenario of Dekel and Birnboim
\citep*{DekelBirnboim06,BirnboimDekelNeistein07,DekelBirnboim08} there
is a transition from a cold accretion regime, in which \lya\ photons
are produced by cooling gas, to a 2-phase medium, where \lya\ may
still be emitted by filamentary `cold flows', and finally to a single
hot phase.  These transitions occur naturally as a result of the
gravitational build-up of structure and the consequent formation of a
virial shock.  It may therefore be possible to follow this progression
via \lya\ imaging surveys, particularly using high spatial resolution
data as we have presented here to study the morphology of the line
emission.  The state of the \lya\ emitting gas would provide direct
clues to the state of the galaxy formation process and the
thermodynamic properties of the gas.  This ``\lya\ calorimeter'' could
be combined with data from the next generation X-ray and mm/radio
telescopes to generate a complete census of the gas in and around
massive galaxies from $z \sim 4$ to $z=0$.

\acknowledgments

We thank the anonymous referee for several helpful comments.  Support
for program \# 9166 was provided by NASA through a grant
(GO-9166.01-A) from the Space Telescope Science Institute, which is
operated by the Association of Universities for Research in Astronomy,
Inc., under NASA contract NAS 5-26555.  AD and MED are supported by
NOAO, which is operated by the Association of Universities for
Research in Astronomy (AURA) under a cooperative agreement with the
National Science Foundation.



\end{document}